# Transport-Weighted Coercivity in Granular $CrO_2$: Why Magnetoresistance and Magnetization Can Disagree

E. Yu. Beliayev[1], I. G. Mirzoiev[1], V. A. Horielyi[1], I. A. Chichibaba[2]

[1] B. Verkin Institute for Low Temperature Physics and Engineering, National Academy of Sciences of Ukraine, Kharkiv, Ukraine.

[2] National Technical University "Kharkiv Polytechnic Institute", Kharkiv, Ukraine.

## Abstract

Magnetoresistance hysteresis is often used to estimate the coercive field of a magnetic material. We reanalyse our published magnetotransport and magnetometry data on compacted powders of the half-metallic ferromagnet $CrO_2$ and show that this identification can fail in a granular conductor. The field $H_p$ at the resistance maximum may be close to, above, or below the bulk coercive field $H_c$, depending on temperature and measuring current. For three samples with different particle shapes and insulating-shell thicknesses, the digitized ratio $\eta = H_p/H_c$ follows the same qualitative sequence and suggests a common crossover near 15 K. We interpret this behaviour as a change in which interparticle junctions control the electrical response. Bulk magnetometry averages over the entire magnetic volume, whereas tunnelling resistance gives the largest weight to the small set of links that carries the current. Cohn's exact network-sensitivity theorem provides a rigorous linear-response definition of these transport weights. In simple terms, Cohn showed that the influence of a given contact on the total resistance is proportional to the square of the current flowing through that contact. Combined with known magnetic pair-correlation effects, this leads to the concept of transport-weighted coercivity. In granular sensors and spin-dependent composites, the resistance peak cannot automatically be treated as the bulk coercive field. Conversely, the difference $H_p - H_c$ may serve as a diagnostic of current-path localization, barrier evolution, and device-to-device variability.

> **Significance.** In a granular magnetic material, the electrical current does not sample every particle equally. At low temperature it may pass through only a few critical contacts. The field read from the resistance loop can therefore describe those contacts rather than the sample as a whole. This is both a warning and an opportunity: it limits the use of magnetoresistance as a direct measure of bulk coercivity, but it also turns the electrical hysteresis into a sensitive probe of the active current path.

## 1. Introduction

In a uniform magnetic conductor, it is natural to expect the resistance and magnetization to switch at approximately the same field. This expectation is often transferred to granular ferromagnets. The field $H_p$ at which the resistance reaches its maximum is then identified with the coercive field $H_c$ obtained from bulk magnetometry.

A granular conductor is different. Its particles are separated by thin insulating barriers, and the current must tunnel from one particle to another. The measured resistance is therefore controlled not only by the magnetic state of the particles, but also by the small subset of contacts that forms the active current path. Two samples with the same bulk magnetization can have different electrical responses if the current follows different paths.

There is a second source of disagreement. Tunnelling depends on the relative orientation of neighbouring magnetic moments, whereas bulk magnetometry measures their average projection along the field. Exchange coupling and randomly oriented anisotropy axes can make these pair correlations evolve differently from the square of the bulk magnetization [1]. Thus magnetization, magnetic pair correlations, and electrical transport are three related but non-equivalent observables.

Compacted $CrO_2$ powders provide a useful test case because their electrical transport combines spin-dependent tunnelling through structurally and chemically nonuniform barriers with strongly percolative current paths [2] [3] [4] [5]. Particle shape, magnetic anisotropy, and barrier thickness and composition further modify the low-temperature resistance and magnetoresistance [6] [7] [8] [9] [10] [11]. In particular, previous measurements revealed that the resistance-peak field $H_p$ does not follow the bulk coercive field $H_c$ in a simple way: $H_p$ approaches $H_c$ at elevated temperature, exceeds it in an intermediate range, and falls below it at the lowest temperatures [11] [12] [13]. The present work places these observations in a single phenomenological framework and discusses their practical meaning.

## 2. Samples and operational definitions

This work presents a new analysis of our previously published measurements on three compacted $CrO_2$ powders. Each specimen is identified by particle morphology and nominal shell thickness. The letter A denotes acicular particles, R denotes rounded particles, and the number gives the shell thickness in nanometres.

| Sample # | Particle morphology | Shell | Thickness | Role in the comparison |
|---|---|---|---|---|
| A-1.73 | Acicular, 22.9×302 nm | β-CrOOH | 1.73 nm | Temperature-dependent $H_p/H_c$; pronounced intermediate maximum |
| R-3.6 | Rounded, ≈120 nm | β-CrOOH | 3.6 nm | Reduced shape anisotropy; temperature- and current-dependent measurements |
| A-2.1 | Acicular, 22.9×302 nm | $Cr_2O_3$ | 2.1 nm | Largest normalized $H_p/H_c$ maximum |

**Table 1. Specimens used in the normalized comparison. A-2.1 corresponds to sample 2 of Ref. [12] and sample C of Ref. [13]; the latter reports additional measurements on the same physical specimen.**

The bulk coercive field $H_c$ is defined by the positive-field zero crossing of the magnetization $M(H)$ on a specified hysteresis branch. The resistance-peak field $H_p$ is defined as the positive-field resistance maximum on the corresponding branch. Their relative displacement is characterized by the ratio

$$\eta(T) = H_p(T)/H_c(T).$$

Values $\eta \approx 1$ indicate agreement between the electrical and magnetic field scales. Values $\eta > 1$ or $\eta < 1$ indicate that the transport-defined field scale lies above or below, respectively, the coercive field obtained from bulk magnetometry.

## 3. Main phenomenology: one sequence in three specimens

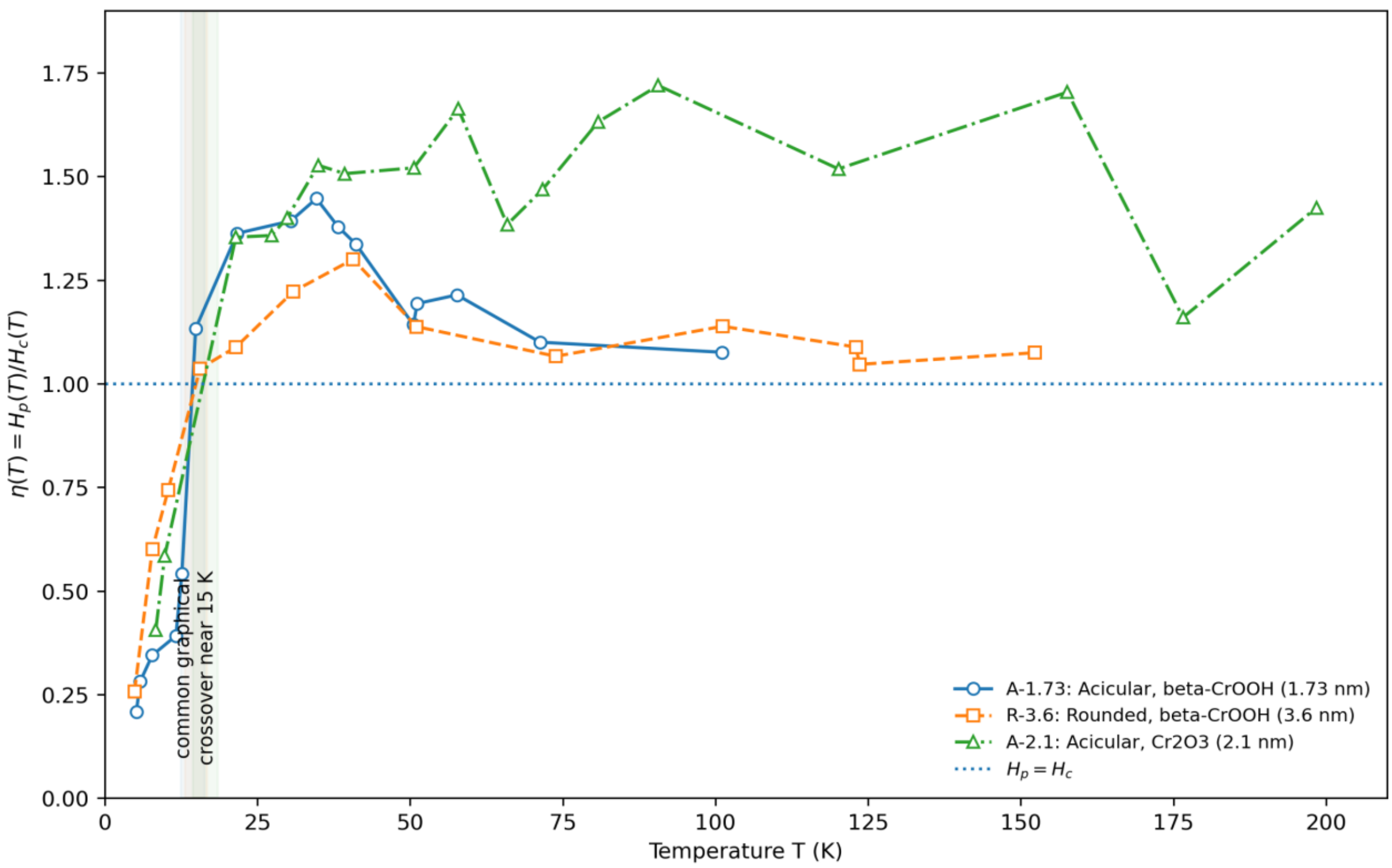

**Fig. 1. Normalized ratio η(T) = $H_p(T)/H_c(T)$ for three compacted $CrO_2$ specimens. The $H_p(T)$ and $H_c(T)$ values were digitized from Fig. 3 of Ref. [13], and $\eta(T)$ was calculated from these values. The vertical shaded ranges show the uncertainty in locating the low-temperature crossings from the digitized curves; they are not experimental confidence intervals. The three datasets suggest a common graphical crossover near 15 K, with baseline estimates spanning approximately 14–16 K.**

Fig. 1 condenses the central experimental observation. At elevated temperature, $H/_p$ is close to $H_c$. On cooling, $H_p$ rises above $H_c$ and reaches a sample-dependent maximum. At still lower temperature, the ratio crosses unity and falls sharply below one. The detailed magnitude and temperature of the maximum depend on particle morphology and barrier properties, but the qualitative sequence is shared by all three specimens.

$$H_p \approx H_c \rightarrow H_p > H_c \rightarrow H_p < H_c. \tag{1}$$

The common crossing near 15 K is an empirical feature of the digitized plots, not evidence for a thermodynamic phase transition. It should therefore be described as a graphical crossover. The important result is the change in sign of $H_p - H_c$: the electrical measurement samples a magnetic population that changes relative to the population measured by bulk magnetometry [12] [13].

## 4. Why the electrical and magnetic fields can disagree

Bulk magnetometry averages over the magnetic moments of all particles, with larger particles contributing more strongly. The condition $M(H_c) = 0$ states that the net moment vanishes. It does not specify how neighbouring moments are oriented relative to one another.

Spin-dependent tunnelling is controlled by the relative orientation of two particles joined by a barrier. Even when the net magnetization is zero, many neighbouring pairs can remain partly correlated. Theory shows that exchange interaction and random anisotropy can make the pair correlation $C_{mag}(H)$ differ from $m^2(H)$, where $m = M/M_s$ [1]. This alone can separate magnetoresistance from magnetization. Exchange coupling can preserve the relative alignment of neighbouring particles after the net magnetization has crossed zero, while random easy-axis orientations broaden their reversal. The pair correlation may therefore reach its characteristic field scale above the bulk coercive field.

The electrical network adds another level of selectivity. A weakly conducting contact that carries almost no current has little effect on the measured resistance, whereas a critical contact on the main current path can dominate the result [14]. For a passive linear two-terminal resistor network, Cohn proved the exact identity [15]

$$\partial R_{\mathrm{net}}/\partial R_{\mathrm{e}} = (I_{\mathrm{e}}/I)^2, \tag{2}$$

where $R_{\mathrm{net}}$ is the measured resistance, $R_e$ is the resistance of one link, I is the total current, and $I_e$ is the current through that link. The corresponding normalized weight is

$$s_{\mathrm{e}} = I_{\mathrm{e}}^2 R_{\mathrm{e}}/(I^2 R_{\mathrm{net}}), \quad \sum_e s_{\mathrm{e}} = 1. \tag{3}$$

This identity is exact for a passive linear network at a fixed operating state. At finite bias, heating and nonlinear tunnel conduction may change both the link resistances and the current distribution.

Thus, the relevant magnetic correlation for transport is not an equal average over all contacts, but the current-weighted average $C_{\mathrm{tr}}(H) = \sum_e s_{\mathrm{e}} \cos\theta_{\mathrm{e}}$. The hierarchy can be summarized as

$$M^2(H) \not\equiv C_{\mathrm{mag}}(H) \not\equiv C_{\mathrm{tr}}(H). \tag{4}$$

The resistance maximum $H_{\mathrm{p}}$ is therefore an operational field scale of the current-carrying network. We call it a transport-weighted coercive field. The term does not imply a new thermodynamic coercivity; it identifies what the electrical experiment actually measures.

## 5. Interpretation of the temperature evolution

At elevated temperature, thermal activation is expected to broaden the set of interparticle links contributing to transport. The electrical measurement samples a relatively broad part of the specimen, so the transport-weighted field approaches the bulk coercive field: $H_{\mathrm{p}} \approx H_{\mathrm{c}}$.

In the intermediate range, $H_{\mathrm{p}} > H_{\mathrm{c}}$. Magnetic pair correlations can continue to evolve after the net magnetization has crossed zero [1]. Electrical disorder may further emphasize contacts associated with more strongly correlated or magnetically harder regions. The resistance peak then occurs after the bulk magnetic zero crossing.

At low temperature, the distribution of effective tunnel resistances is expected to broaden, concentrating the current in fewer critical contacts. Strong-disorder network studies show that a small number of links, or even one link, can control the macroscopic resistance [16]. If the surviving path contains electrically favourable but magnetically softer particles or multidomain bottlenecks, those links can reverse before the bulk sample and produce $H_{\mathrm{p}} < H_{\mathrm{c}}$. More generally, the active path may preferentially sample particles or junctions with smaller anisotropy barriers or lower local switching fields. This interpretation is deliberately phenomenological. The active path was not imaged directly, and the magnetic character of its critical particles was not measured independently. The data are consistent with a change in transport selection; the proposed soft-bottleneck picture is the simplest mechanism consistent with the low-temperature trend.

## 6. Measuring current reconfigures the readout

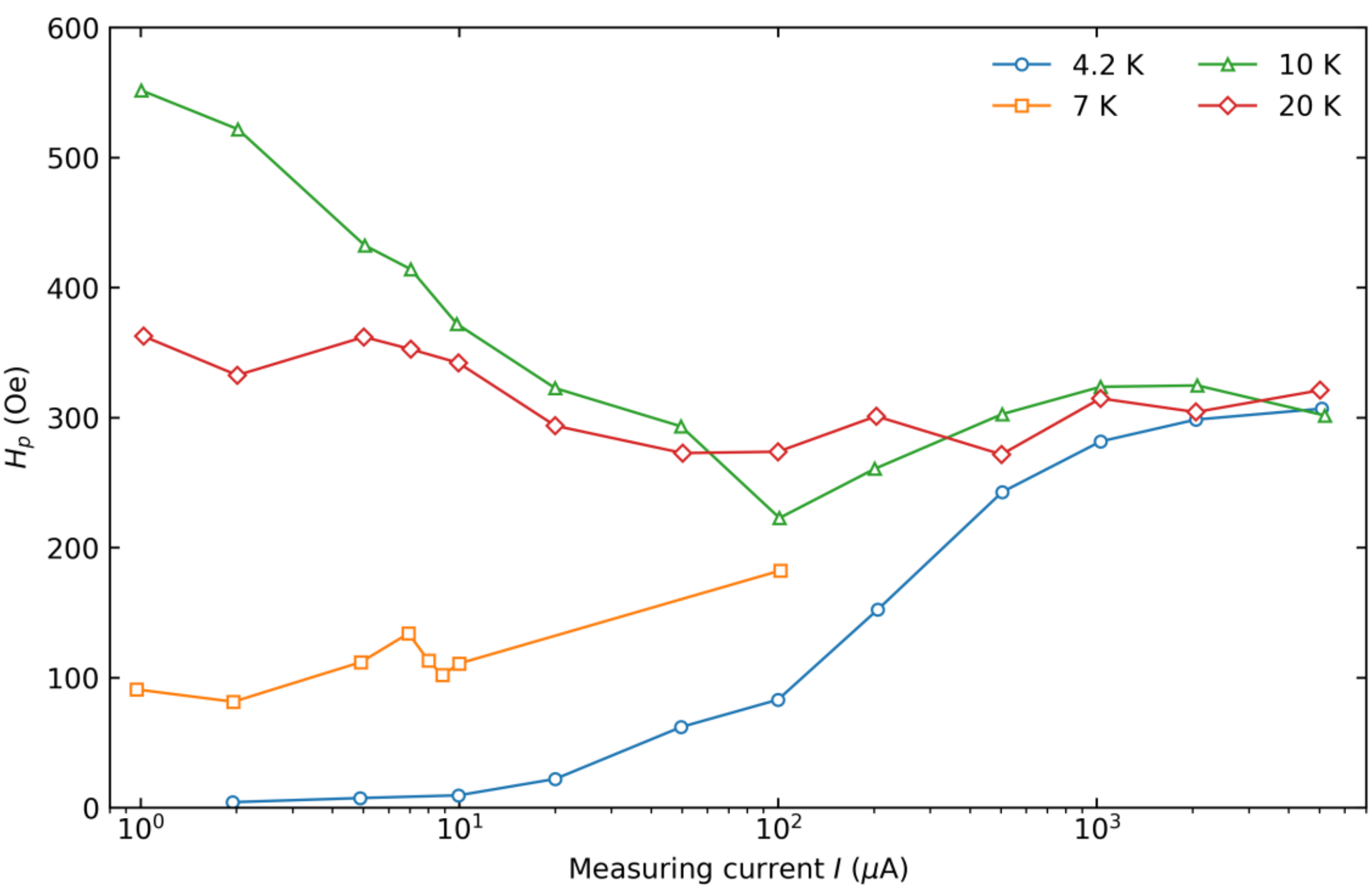

**Fig. 2. Field $H_p$ of the magnetoresistance maximum versus measuring current for specimen R-3.6. The temperature assignment follows the published caption: circles, 4.2 K; squares, 7 K; triangles, 10 K; diamonds, 20 K. Data were digitized from Fig. 7 of Ref. [13]. Lines are guides to the eye.**

The current dependence confirms that $H_p$ is not a fixed property of the bulk magnet alone. At 20 K, $H_p$ remains close to the bulk coercive-field scale. At 10 K it first decreases, consistent with local heating of the most resistive links, and then returns toward approximately 300 Oe as additional branches become electrically active. At 4.2 and 7 K, $H_p$ rises from a low-current value toward the same bulk scale. The simplest common interpretation is that increasing bias changes the active current path. The current therefore acts both as a probe and as a perturbation. The finite-bias data are used as phenomenological evidence of network reconfiguration, rather than as a direct application of the linear identity in Eq. (2). Related low-temperature measurements on Fe-containing $CrO_2$ powders showed that the shape of the MR hysteresis also depends on the magnetic-field sweep rate, demonstrating the role of spin relaxation and observation time [17]. An influence of spin-polarized current on domain walls remains possible at the lowest temperatures, but the existing data do not separate it uniquely from Joule heating and voltage-assisted recruitment of additional links.

## 7. Practical implications

*Sensor calibration.* A magnetoresistance peak should not automatically be used as a measure of the bulk coercive field in a granular sensor. The apparent switching field can depend on temperature, measuring current, contact geometry, and magnetic history even when the bulk magnetization loop is unchanged.

*Device design.* The electrical switching field may, in principle, be tuned separately from the bulk magnetic coercivity by controlling particle morphology, barrier thickness, barrier uniformity, and bias. This provides a design route for granular spin-dependent composites and resistive magnetic elements, although the present work does not demonstrate an optimized device.

*Network diagnostics.* The difference $H_p - H_c$ may serve as a diagnostic of current-path localization, barrier evolution, and device-to-device variability. A growing discrepancy may indicate stronger current localization, barrier degradation, contact ageing, or current crowding. Multi-contact measurements could use this sensitivity to identify spatially unstable regions in a composite.

The main applied message is therefore a measurement principle: the resistance loop reports the magnetic state of the electrically active network, not necessarily the magnetic state of the whole specimen.

## 8. Limitations and direct experimental tests

- The replotted data in Figs. 1 and 2 were digitized from published graphs. The uncertainty shown in Fig. 1 describes graphical extraction, not pointwise experimental error.
- The temperature range around 15 K is a common crossover in the available datasets, not a universal critical temperature.
- Exchange-induced pair correlations and random easy-axis anisotropy can separate magnetoresistance from magnetization even without strong transport selection; the present data do not uniquely decompose these contributions.
- The active current path has not been imaged directly. The proposed role of soft multidomain bottlenecks remains an interpretation.
- Cohn's identity applies exactly only in linear response. Strongly nonlinear current-dependent measurements require a local differential treatment.
- Multi-contact measurements, low-frequency resistance noise, repeated magnetic cycling, and pulsed-current experiments provide direct tests of the transport-selection picture.

## 9. Conclusions

The field of a magnetoresistance maximum in granular $CrO_2$ is not generally identical to the bulk coercive field. Three distinct specimens show the same qualitative temperature sequence: $H_p \approx H_c$, then $H_p > H_c$, and finally $H_p < H_c$. The measuring current also shifts $H_p$ and can drive it back toward the bulk field scale.

These observations follow naturally once three levels of magnetic information are separated: bulk magnetization, magnetic pair correlations, and the current-weighted correlations of the active tunnelling network. Cohn's sensitivity identity gives an exact linear-response meaning to the transport weights and supports the operational concept of transport-weighted coercivity.

The applied consequence is straightforward. In a granular magnetoresistive material, the electrical hysteresis is a readout of the current path as well as of the magnet. This must be accounted for when calibrating sensors or interpreting switching fields, and it may be exploited to monitor current localization and barrier stability.